\documentclass[aps,twocolumn,groupedaddress,prb,floatfix,showpacs]{revtex4}
\usepackage{graphicx}
\usepackage{dcolumn}
\usepackage{bm}
\usepackage{amsmath}
\usepackage{verbatim}

\begin{document}

\title{Intrinsic and extrinsic perturbations on the topological insulator Bi$_2$Se$_3$ surface states}

\author{Jiwon Chang}
\email{jiwon.chang@mail.utexas.edu}
\author{Priyamvada Jadaun}
\author{Leonard F. Register}
\author{Sanjay K. Banerjee}
\author{Bhagawan Sahu}
\affiliation{
Microelectronics Research Center, The University of Texas at Austin, Austin Texas 78758
}

\date{\today}

\begin{abstract}
Using a density functional based electronic structure method, we study the effect of perturbations on the surface state Dirac cone of a strong topological insulator Bi$_2$Se$_3$ from both the intrinsic and extrinsic sources. We consider atomic relaxations, and film thickness as intrinsic and interfacial thin dielectric films as an extrinsic source of perturbation to the surface states. We find that atomic relaxations has no effect on the degeneracy of the Dirac cone whereas film thickness has considerable effect on the surface states inducing a gap which increases monotonically with decrease in film thickness. We consider two insulating substrates BN and quartz as dielectric films and show that surface terminations of quartz with or without passivation plays critical role in preserving Dirac cone degeneracy whereas BN is more inert to the TI surface states. The relative orbital contribution with respect to bulk is mapped out using a simple algorithm, and with the help of it we demonstrate the bulk band inversion when spin-orbit coupling is switched on. The layer projected charge density distributions of the surface states shows that these states are not strictly confined to the surface. The spatial confinement of these states extends upto two to three quintuple layers, a quintuple layer consists of five atomic layers of Bi and Se. 
\end{abstract}

\maketitle

\section{Introduction}

Three dimensional (3D) topological band insulators (TI), Bi$_2$X$_3$, X=Se, Te and Sb$_2$Te$_3$ have attracted considerable attention from the condensed matter physics community because of the relatively simple crystal structure that hosts its novel surface states\cite{zhang1}. Many more 3D TI materials have been predicted\cite{new3d} by now and quest for studying their novel surface state properties has increased dramatically. The surface state of 3D TI are time-reversal symmetric (TRS) in some regions of momentum space and are therefore protected against perturbations which cannot break TRS. These considerations also suggest the possibility of dissipationless transport within these surface states where it would be relatively insensitive to non-magnetic disorder or any perturbations that protect the TRS properties of these states\cite{zahid}. Such properties have caused excitement in the electron device community. The advances in undertanding of structural, electronic, magnetic and transport properties of 3D TI, made possible by both experimental and theoretical studies\cite{zhang2}, have allowed it to gain attention in the scientific community\cite{star1}. It is, however, not clear whether perturbations, both intrinsic and extrinsic, can affect the surface state Dirac cone, by breaking the TRS. Therefore, it is necessary to study 3D TI and the novel surfaces states, using {\it ab-initio} methods. To understand the effects of these variables on the surface states, we report on the density functional based electronic structure studies of one such strong 3D TI Bi$_2$Se$_3$. Our study suggests a critical thickness of six quintuple layers (QLs) needed to maintain the degeneracy of the electron-hole bands at the Dirac point and the surface states have spatial extension to within 2$\sim$3QLs. Atomic relaxations have no effect on the linear spectrum but thin dielectric films (crystalline BN and SiO$_2$) are found to play a critical role in breaking the surface state degeneracy, depending upon the nature of surface terminations and passivations. Our studies will be useful for designing electronic devices using a 3D TI and will have implications in interpreting experiments.  

We begin by describing the bulk crystal structure of Bi$_2$Se$_3$ and the computational method used for this study in section II. In section III, we present the electronic structure of bulk TI and a simple algorithm to map out the surface states from the bulk band structure, and we use this method to explain band inversion when spin-orbit coupling is switched on. We then discuss the thin-film structure and the thickness dependent surface band structure of Bi$_2$Se$_3$ in section IV. The spatial extension of these surface states are then determined using charge densities projected onto each atomic layer and each atom. The effect of atomic relaxations and thin dielectric films on the surface electronic structure will be discussed in section V. Finally we present our summary and conclusions.

\begin{figure}[ht!]
\scalebox{0.255}{\includegraphics{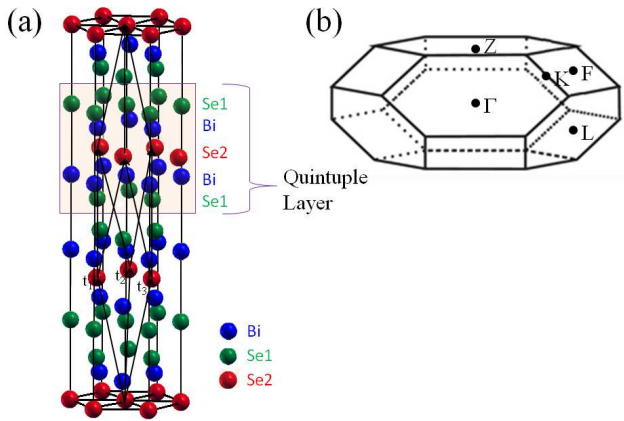}}
\caption{ (Color online) (a) Hexagonal crystal structure of bulk Bi$_{2}$Se$_{3}$ derived from the trigonal structure whose three primitive vectors t$_1$, t$_2$ and t$_3$ are indicated. The trigonal structure is part of the larger hexagonal structure. The hexagonal cell contains three times the number of atoms compared to that in the trigonal structure and the atomic layers of Bi (blue), Se1 (green) and Se2(red) are stacked along the {\it z}-direction. One quintuple layer containing five atomic layers is indicated in a square region. Se1 and Se2 are two inequivalent sites of the Se atoms in the unit cell (b) The first BZ of the bulk trigonal structure with four time-reversal invariant points  $\Gamma$, Z, F and L are shown.
}
\label{fig:Fig1}
\end{figure}

\section{Bulk Structure and Computational Method}

The bulk crystal structure of Bi$_2$Se$_3$ is trigonal (or rhombohedral) with lattice constant of {\it a}= 0.984 nm and $\alpha$$\sim$25$^o$. The primitive cell consists of five atoms (two Bi and three Se atoms) arranged in a order Se2-Se1-Bi-Bi-Se1 where Se1 and Se2 are two inequivalent sites of Se atoms in the cell. Figure 1(a) shows a cartoon of bulk rhombohedral structure along with three primitive lattice vectors that span it (the tracing inside the large cell). The three-fold rotation axis is taken along the {\it z}-axis. Alternatively, one can construct a hexagonal cell from the trigonal cell so that the number of atoms is tripled with the lattice parameters {\it a}=0.4138 nm and {\it c}=2.8633 nm (Fig. 1(a), the larger cell). The building block of the hexagonal bulk Bi$_2$Se$_3$ crystal consists of five atomic layers, each layer containing only one atom. One such building block is referred to as quintuple layer (QL) (square shaded region in Fig. 1(a)). The hexagonal unit cell contains three such QLs, i.e., 15 atomic layers stacked along the {\it z}-direction. The atomic planes are arranged in a sequence Se1-Bi-Se2-Bi-Se1, which is different from the trigonal case, with the following interlayer distances: {\it d}(Se1-Bi)= 0.156344 nm, {\it d}(Bi-Se2)=0.190665 nm, {\it d}(Se2-Bi)=0.190665 nm, {\it d}(Bi-Se1)=0.156353 nm and {\it d}(Se1-Se1')=0.259304 nm where Se1' is the Se1 site in the next unit cell. Because of the relatively larger distance between two QLs, their anchoring along the {\it z}-direction is considered to be rather weak. Indeed, mechanical exfoliation of Bi$_2$Se$_3$ demonstrated recently\cite{balan1} hints at van der Waal's type of bonding between two neighboring unit cells, much like a graphene exfoliation.

We use a DFT-based electronic structure method implemented with projector-augmented wave basis and pseudopotentials\cite{vasp1}. Since the topological features of the surface states arise from spin-orbit coupled states of the bulk, we invoke its spin-orbit coupling (SoC) feature\cite{vasp2} and we choose exchange-correlation as Perdew-Burke-Ernzerof (PBE) type\cite{perdew}. The choice of the PBE over local-density approximation is guided by its use in previous structural studies on bulk Bi$_2$Se$_3$ which reproduced the photoemission spectra very well\cite{zhang2}. These studies find the optimized lattice constant of Bi$_2$Se$_3$ close to the experimental value, therefore we did not optimize lattice constants of the bulk structure in our calculations. A kinetic energy cut-off of 25 Ry and a {\bf k}-point mesh of 10 $\times$ 10 $\times$ 10 in the first Brillouin Zone (BZ) (Fig. 1(b)) are used for self-consistent steps as well as generating the bulk band structures. Larger energy cut-offs and {\bf k}-mesh size were tested for total energy convergence. The first BZ of the bulk structure contains four time-reversal invariant (TRI) points, namely, $\bf \Gamma$, {\bf L}, {\bf Z} and {\bf F} (Fig.1(b)).

\section{Bulk band structure}

In this section, we address the band structure of bulk Bi$_2$Se$_3$, with and without SoC. We demonstrate the band inversion with switching on of SoC with the help of an algorithm which maps out orbital contribution to the electronic states from bulk bands. Figure 2(a) and (b), respectively, shows energy dispersion curves for the bulk Bi$_2$Se$_3$ without and with SoC, along high-symmetry directions connecting the four TRI points (Fig. 1(b). The direct band gap, in both cases, forms at the $\Gamma$-point and is 200 meV(285 meV) without(with) SoC. These gaps compare well with those obtained in other DFT simulations\cite{zhang2}. It was suggested theoretically, using  a model Hamiltonian based on atomic orbitals and without SoC, that the valence band maximum (VBM) and conduction band minimum (CBM) mainly consists of Se  and Bi $p_z$ orbitals\cite{zhang3}, respectively (Fig. 2(a)). Switching on the SoC inverts the bands at the $\bf \Gamma$-point (Fig. 2(b)) and consequently the contributions from Bi and Se ions associated with these bands.

\begin{figure}[!]
\scalebox{0.27}{\includegraphics[angle=-90]{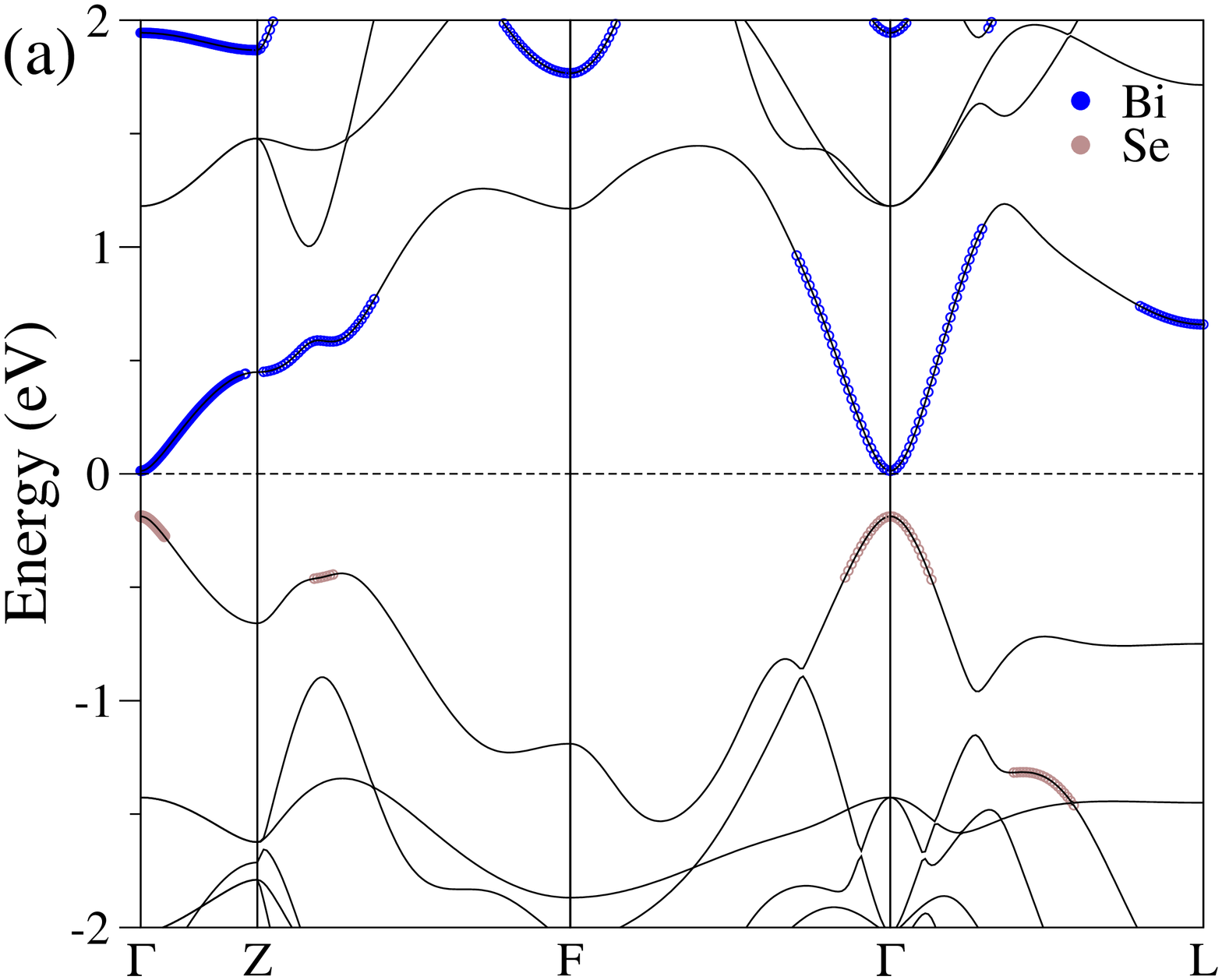}}\\
\scalebox{0.27}{\includegraphics[angle=-90]{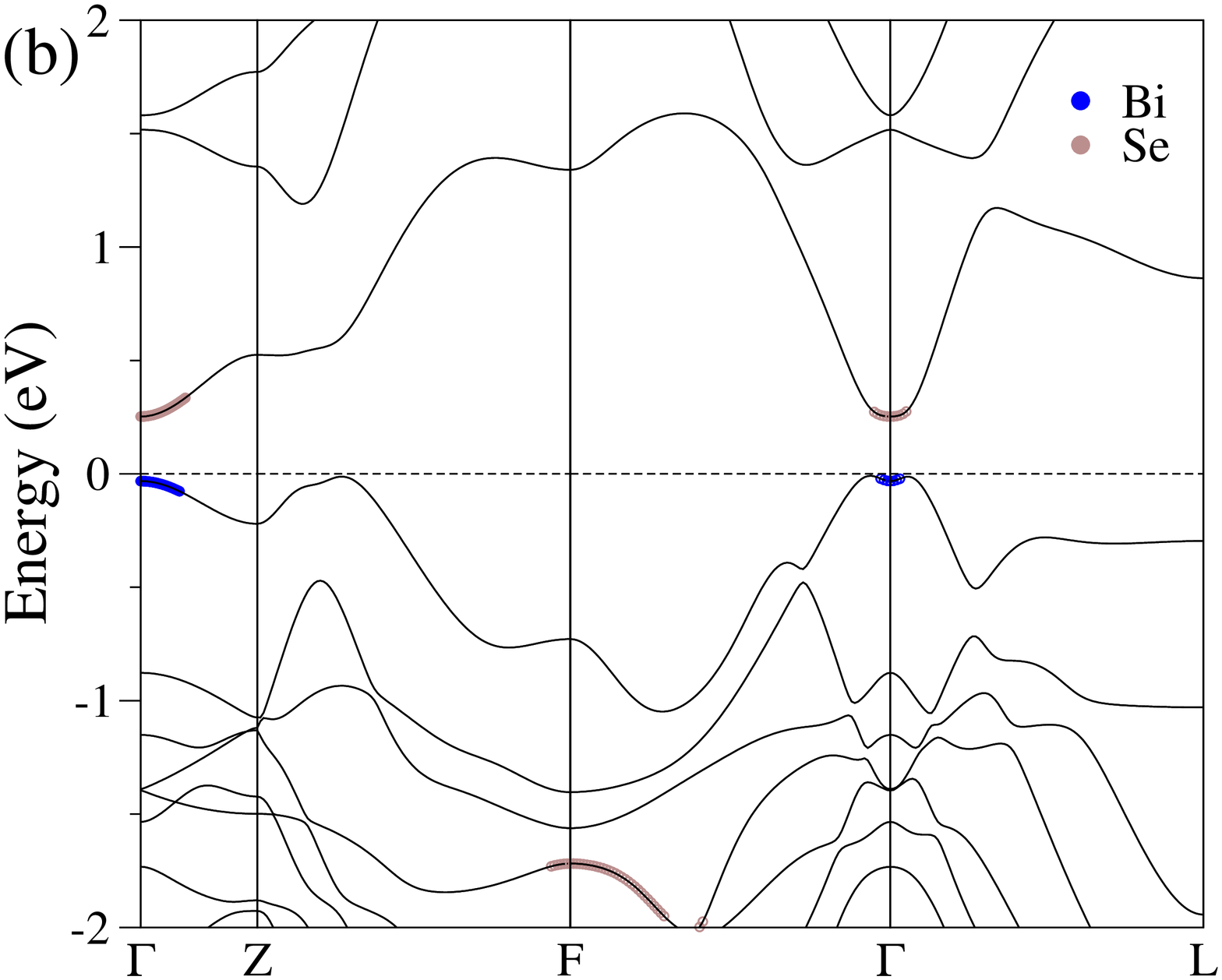}}
\caption{ (Color online) Band structure of trigonal Bi$_{2}$Se$_{3}$ along high symmetry directions in the first BZ connecting the four time-reversal invariant points (a) without spin-orbit coupling and (b) with spin-orbit coupling. Electron states mainly from $p_z$ orbitals of Bi and Se are marked on the top of bulk bands and band inversion process in (b) is clearly seen.}
\label{fig:Fig2}
\end{figure}

To understand the band inversion at the $\Gamma$-point visually, we adopt a simple algorithm to map out the most relevant orbital contribution to VBM and CBM\cite{park}. Since $p_z$ orbitals mainly contribute to these bands near the Fermi level, the algorithm projects the crystal wave-functions on to the angular momentum dependent spherical harmonics at a particular {\bf k}-point with  the associated band energies for each atom. The primitive unit cell of bulk Bi$_2$Se$_3$ contains three Se and two Bi atoms and using the above algorithm, we obtain {\it s} and three {\it p} orbitals contribution for each of the atom type at each {\it k-point} and band energy. We take the ratio of Se and Bi $p_z$ orbitals contributions compared to the total orbital contributions resulting from all the orbitals of five atoms at a particular {\bf k}-point and the band energy. If this ratio is 40$\%$ (60$\%$) or greater for Bi (Se), we take the $p_z$ orbital contribution to that band from Bi or Se at that {\bf k}-point with the associated band energy; otherwise, we neglect it. The choice of the percentage cut-off is guided by the number of each type of atoms in the unit cell: two for Bi versus three for Se. We find that changing the cut-off or critical percentage ratio makes the contributions to each band at each {\bf k}-point different but the overall contribution to the VBM and CBM at the $\Gamma$ point remains the same. In this sense, the choice of critical percentage is arbitrary. We use the above algorithm both without and with SoC calculations, and as a result we could see the band inversion as seen in Fig. 2(b). In the literature, this inversion process is termed as phase transition from a trivial insulator (Bi$_2$Se$_3$ without SoC) to a topological insulator.    

\begin{figure}
\scalebox{0.390}{\includegraphics{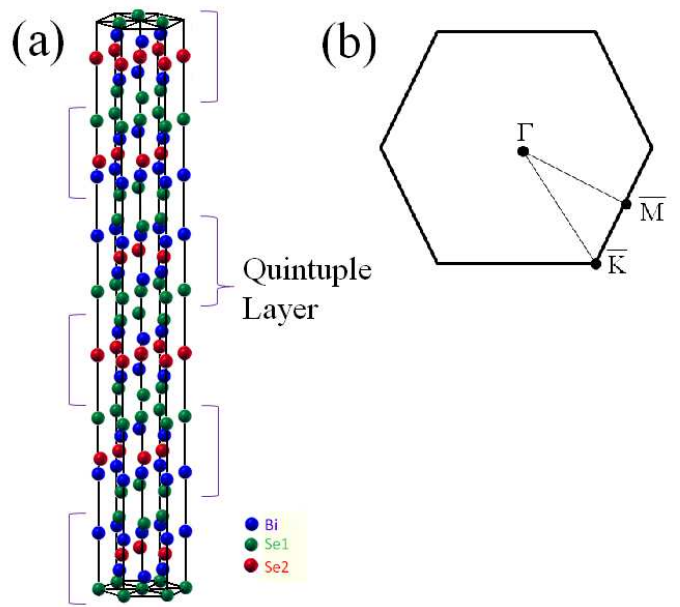}}
\caption{ (Color online) (a) Schematic diagram of 6QL thin film structure obtained by stacking six quintuple layers along 
{\it z}-direction and (b) Two-dimensional Brillouin zone of the (111) surface of the bulk Bi$_2$Se$_3$ with three time-reversal invariant points  $\bar{\Gamma}$, \={M}, and \={K} denoted.
}
\label{fig:Fig3}
\end{figure}

\begin{figure}
\scalebox{0.215}{\includegraphics[angle=-90]{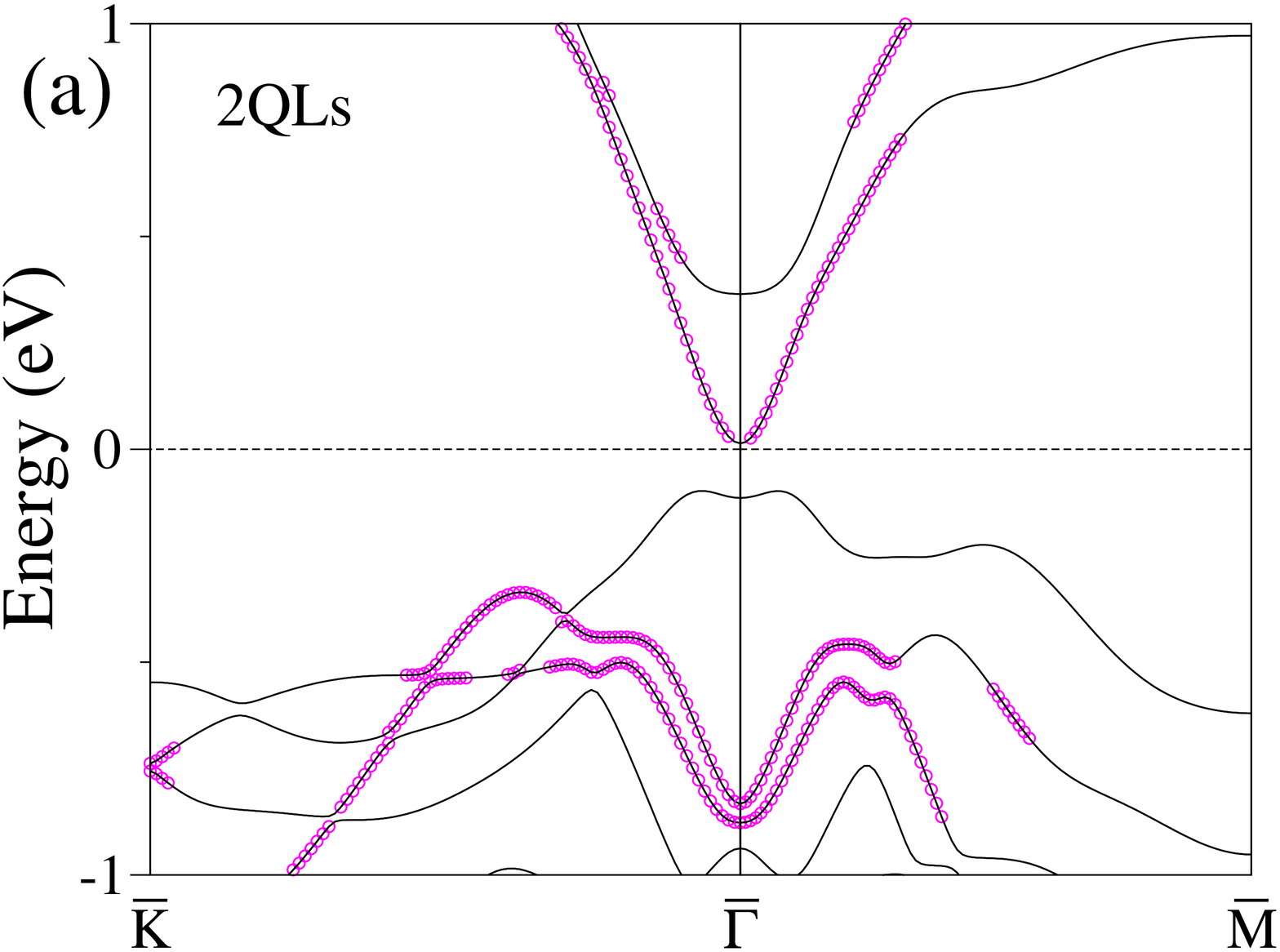}}
\scalebox{0.215}{\includegraphics[angle=-90]{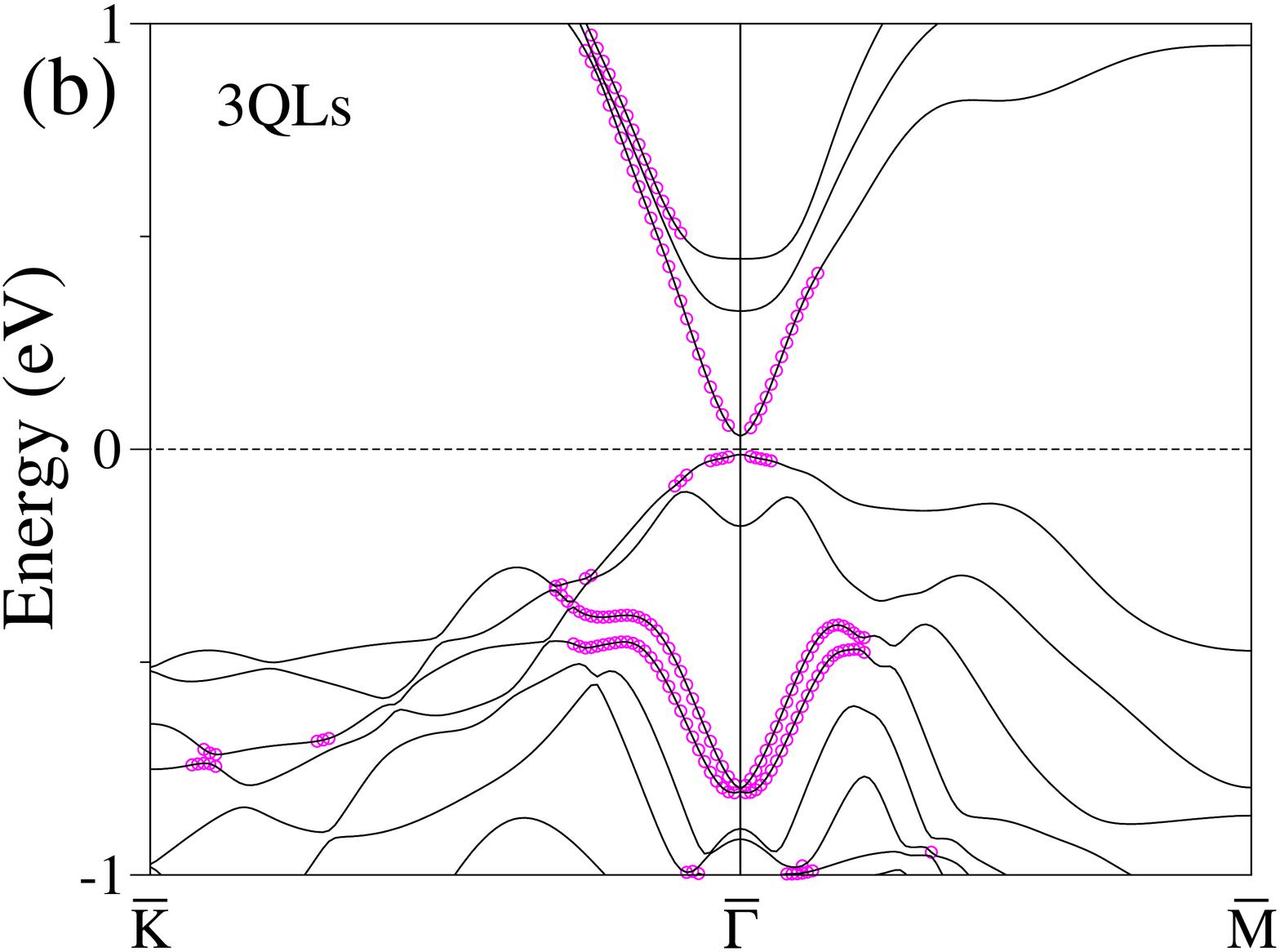}}
\scalebox{0.215}{\includegraphics[angle=-90]{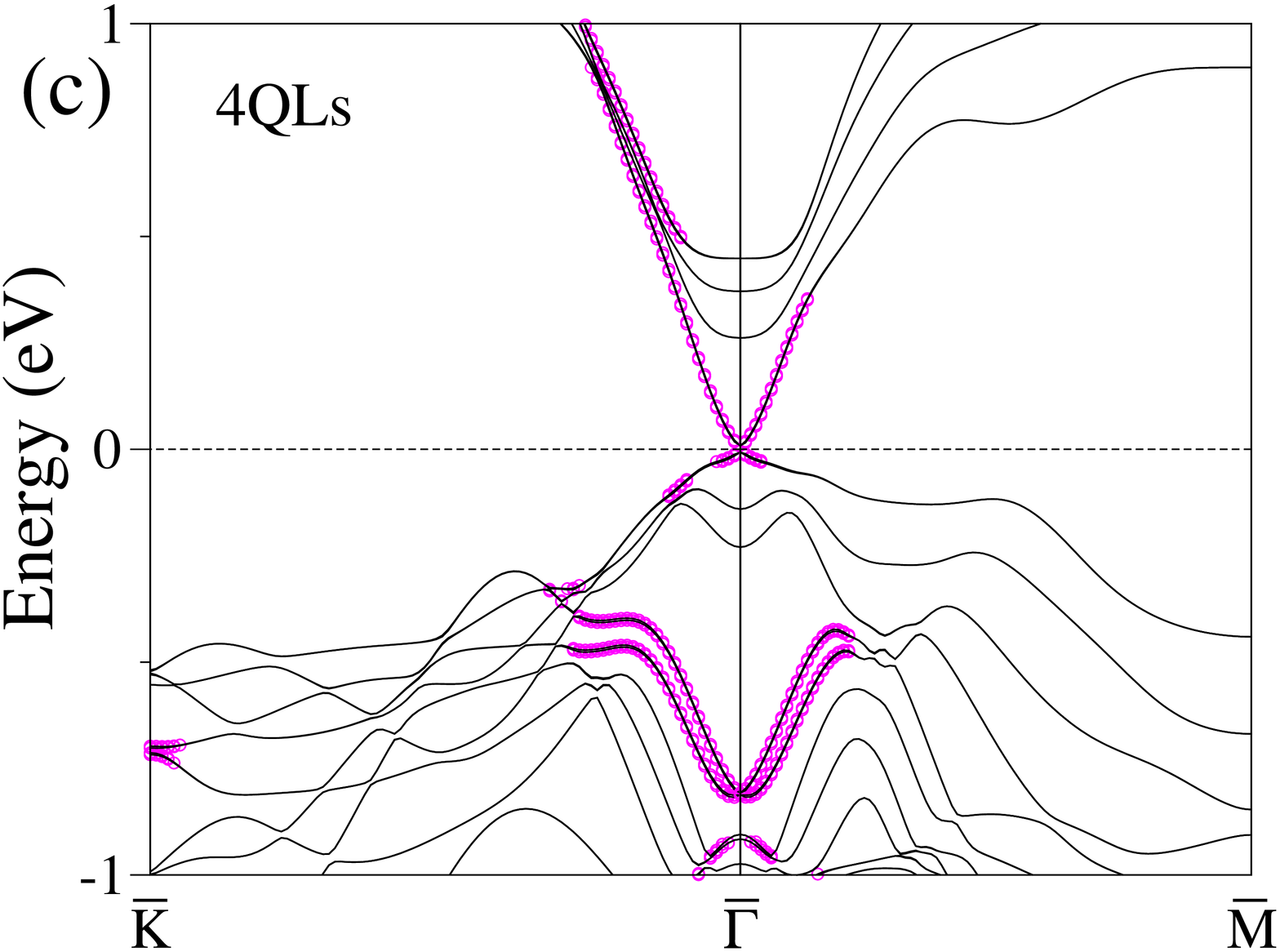}}\\
\scalebox{0.215}{\includegraphics[angle=-90]{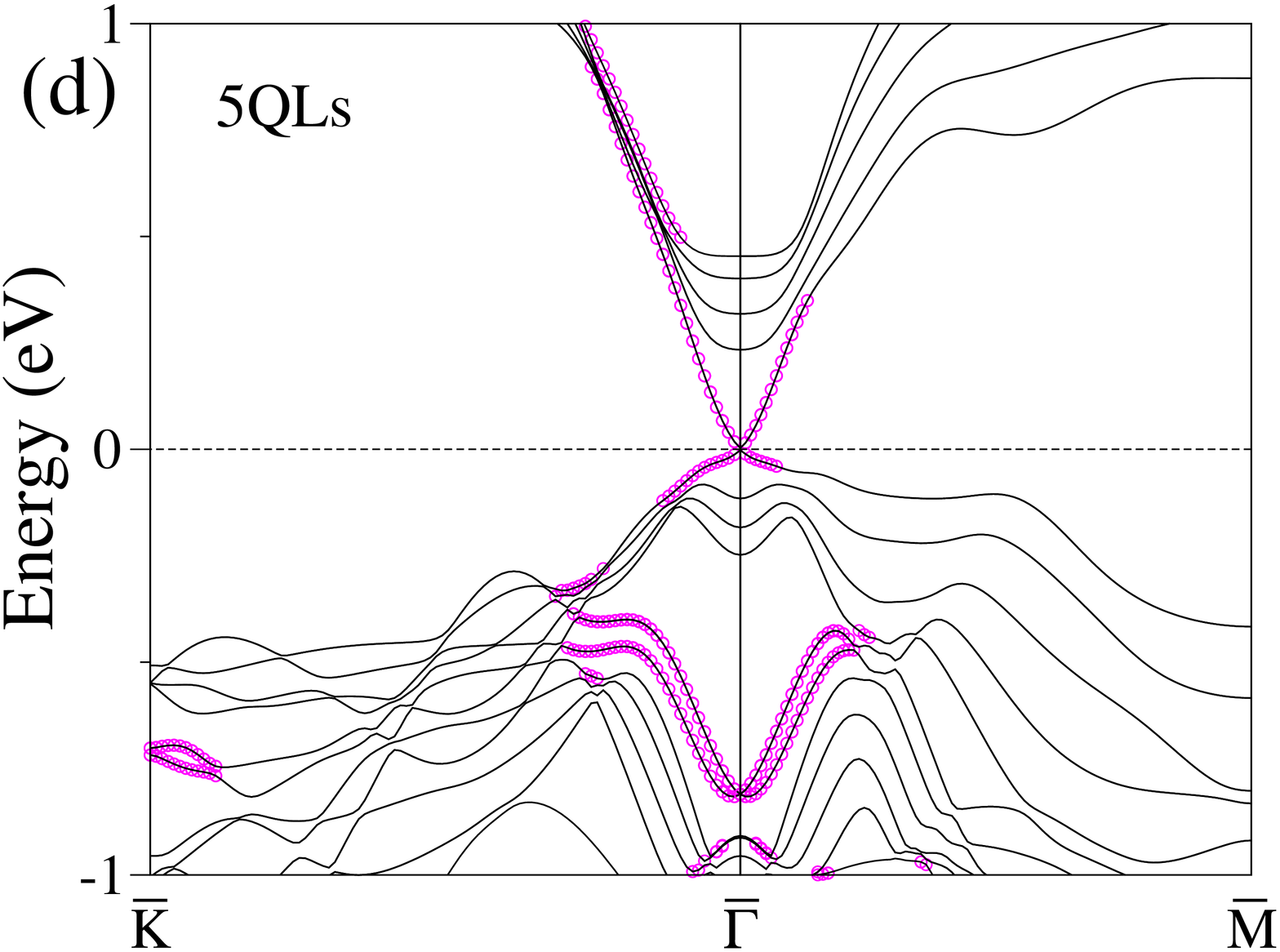}}
\scalebox{0.215}{\includegraphics[angle=-90]{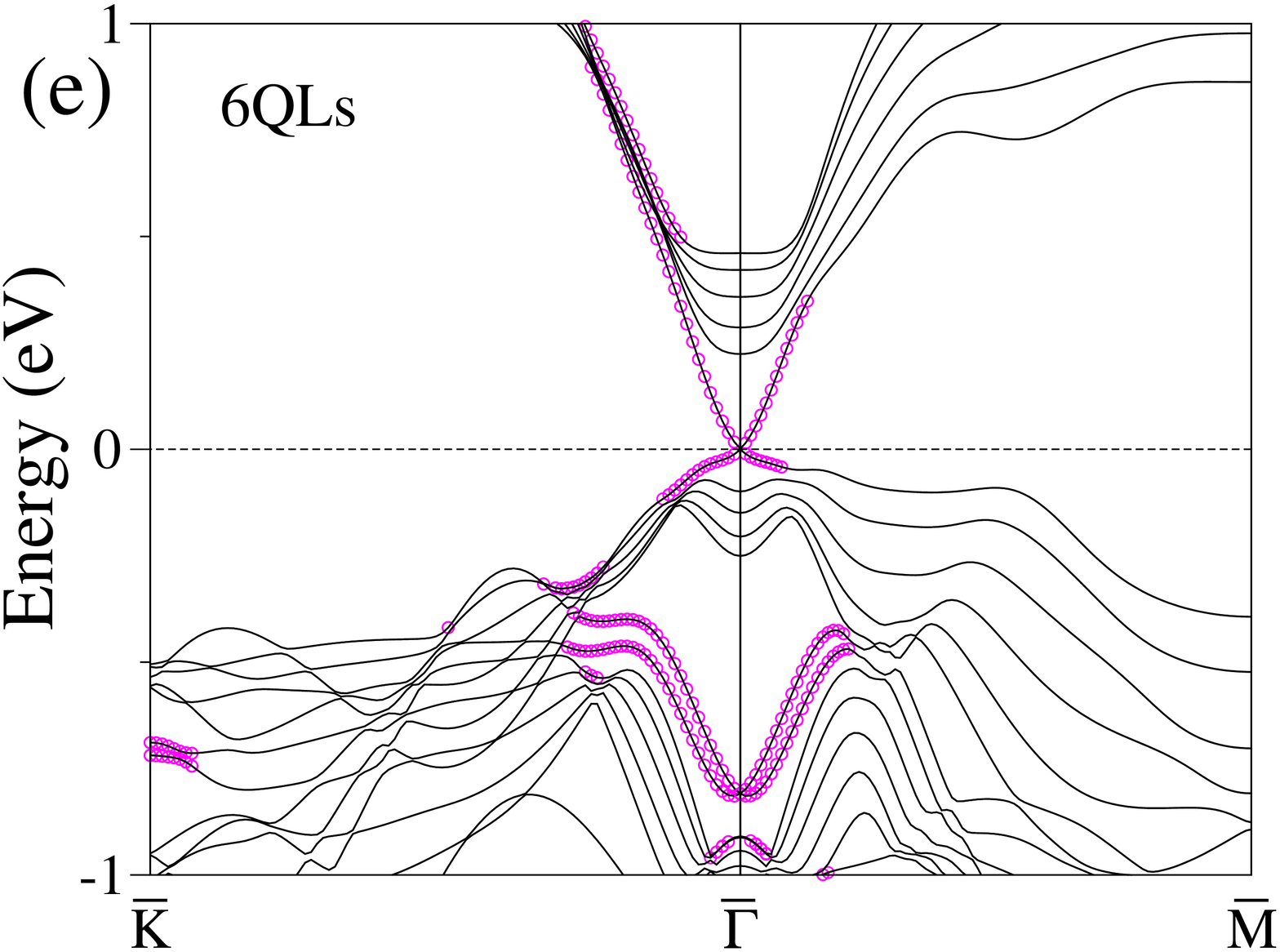}}
\caption{ (Color online) Band structures of Bi$_2$Se$_3$ films with different thicknesses (a) 2QLs (b) 3QLs (c) 4QLs (d) 5QLs and (e) 6QLs. Orbital contributions from first few layers in each thicknesses is marked with red circles. The induction of a gap in thinner structures is clearly seen.}
\label{fig:Fig4}
\end{figure}

\section{Thin film band structures and surface state localization}

The most interesting feature emerges when thin-film structures are considered. A gapless linear spectrum with continuum in the bulk is evident at the TRI point $\Gamma$ (for example Figure 4(e) for a thin film consisting of 6QLs). These states are contributed by both the surfaces of the thin film and are spin-polarized. The thin film structure is constructed by stacking up the quintuple layers along the {\it z}-direction with a vacuum region that forms the supercell in the DFT calculation (Fig. 3(a)). We consider several thicknesses of the thin film structure: 2QLs, 3QLs, 4QLs, 5QLs and 6QLs to study the robustness of the linear energy dispersion and for each thickness we used a vacuum size of about 3 nm. We did not relax the atomic positions. The relaxations and their effect will be considered later. Kinetic energy cut-off of 25 Ry and 10 $\times$ 10 $\times$ 1 {\bf k}-point mesh in the surface BZ of the hexagonal Bi$_2$Se$_3$ (Fig. 3(b)) is used. On decreasing the thickness from 6QLs to 5QLs, a gap seems to be induced at the degenerate point of the linear spectrum at $\Gamma$ (Fig. 4(d)). The magnitude of the induced gap increases monotonically going from 5QLs to 4QLs to 3QLs and finally to 2QLs (Figs. 4(a-c)). These gaps are tabulated in Table I. This suggests a critical thickness of at least 6 QLs in experiments in order to access the novel surface states of Bi$_2$Se$_3$. We note that thickness dependent gaps in Bi$_2$Te$_3$ were realized experimentally\cite{Li} and predicted theoretically\cite{park}. In Bi$_2$Se$_3$, we are aware of only one theoretical work which suggests thickness dependent electronic structure\cite{liu}.     

To understand the origin of the induced gap in the linear spectrum, we first map out all the orbital contributions to the bulk band structure from all the atoms in first few layers of each thickness, from the top and bottom region of the thin film (Figs. 4(a)-(e)). For thickness of 3QLs-5QLs, the contributions from the top or bottom one quintuple layer is chosen whereas for 2QLs system, we consider the top or bottom two atomic layers for estimating the contributions. The contributions are calculated using the same procedure as explained in Section III with two exceptions: 1) The sum of contributions from all the orbitals and atom types are considered (not just from the $p_z$ orbitals) and 2) the critical percentages, for accepting it as a contribution, are different for different thicknesses. The critical percentage for 2QLs is fixed at 30 $\%$ whereas for 3QLs-6QLs, percentages are respectivley, 60$\%$, 55$\%$, 50$\%$ and 45$\%$. The choice of these critical percentages are arbitrary. We find that choosing different percentages do not change the overall features in the band structures. We note that thinner structures have larger critical percentage for accepting the contributions with reduction in percentages for thicker structures. 
This is motivated by the fact that thicker structures have more QLs than the thinner ones and we fixed the contributions to only 1 QL for all thicknesses, except for 2 QL structures where two atomic layers are considered to be a cut-off limit.
 So keeping the same critical percentage for all thicknesses will offset the number of quintuple layers to be considered for acceptance as a contribution and different thickness will need different number of quintuple layers. Alternatively, one can fix the critical percentage for all the thicknessess to the same value and let the number of quintuple layers that contribute to the bulk band structure differ from one thickness to another. Either consideration leads to the same conclusion, namely the contribution to the Dirac cone near $\Gamma$ originates from the first few layers of the thin film.      

To locate the critical number of quintuple layers upto which surface states contributing to the Dirac cone is spread in real space, we use a method in which atomic layer dependent charge density from the surface wave functions is calculated. We choose an energy window of 50 meV around the Fermi level in 4QLs-6QLs bandstructures (Figs 4(c)-(e)) with the states in the neighborhood of the $\Gamma$-point and plot the layer projected relative charge density (Figs. 5(a)-(c)). The relative charge density is defined to be the ratio of maximum charge density in a given layer to the maximum among the maximal charge densities of all the layers in that structure. Our calculations hint at 2-3 QLs spatial extension of the surface states in 4QLs-6QLs. The charge density distribution for 2QLs and 3QLs is not shown because within the energy window of 50 meV considered for this calculations, bulk as well surface states both contribute and it is not a pure surface state contribution (Figs. 4(a) and (b)).       
The surface state localization length to within 2QLs-3QLs in real space hint at possible interactions due to their overlap in thinner structures. This surface state interactions result in opening of the band gap. However, the time-reversal symmetry is not broken at the $\Gamma$-point; the four-fold degenerate band splits into two-fold degenerate bands. The overlap in thinner structures results in finite charge density in a region where surface states for thicker structures had no densities (Figs. 5(a) and (b)). As the degree of overlap increases with decreasing thickness, the induced gap increases monotonically, as seen from Table I. The surface state localization deep inside the bulk region has implications for experiments where measured charge mobility can only be assigned to both suface and bulk states and it will be a challenge in experiments to separate these two contributions.  

\begin{table}
\caption{ Induced band gaps (in eV) for thin-films with various thicknesses at the time-reversal invariant point $\bar{\Gamma}$.}
\begin{tabular}{ c | c | c | c | c | c  }
\hline \hline
  Thickness (QL$\sim$1nm) & 2 & 3 & 4 & 5 & 6 \\
\hline
  Band gap  & 0.1281 & 0.0430 & 0.0155 & 0.0051 & 0.0000\\
at $\bar{\Gamma}$ & & & & &\\
\hline \hline
\end{tabular}
\end{table}

\begin{figure}
\scalebox{0.315}{\includegraphics{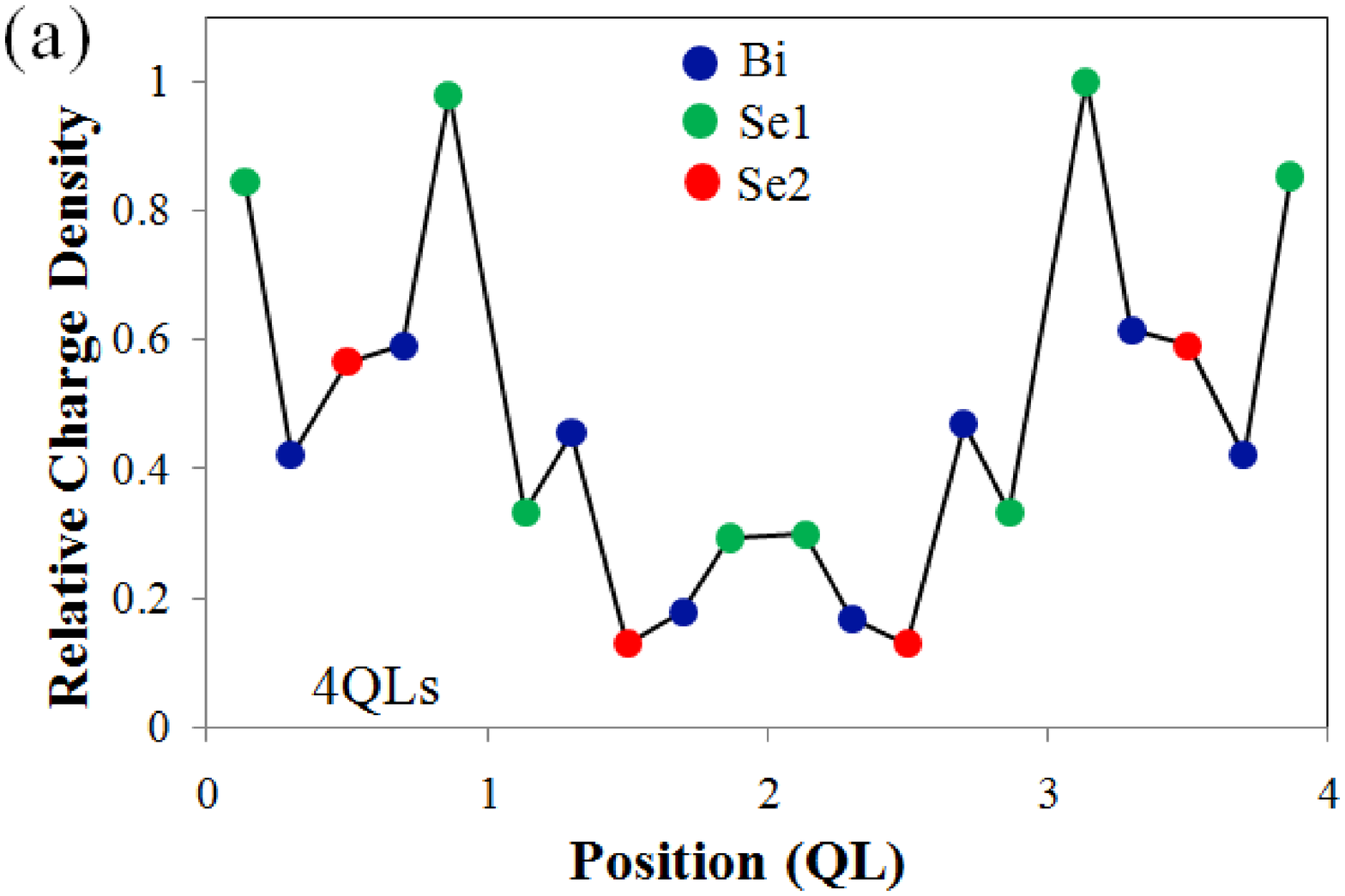}}
\scalebox{0.315}{\includegraphics{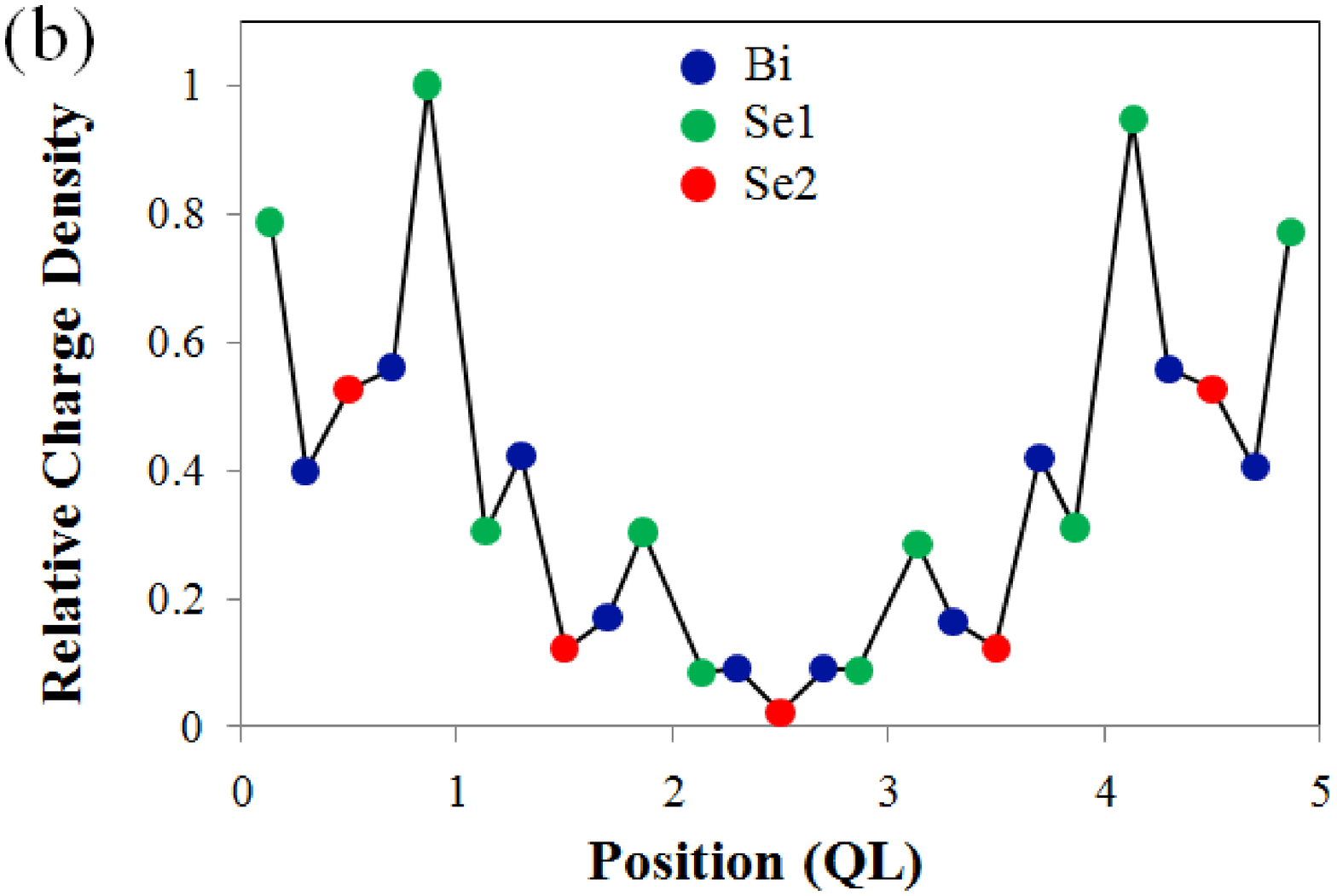}}
\scalebox{0.315}{\includegraphics{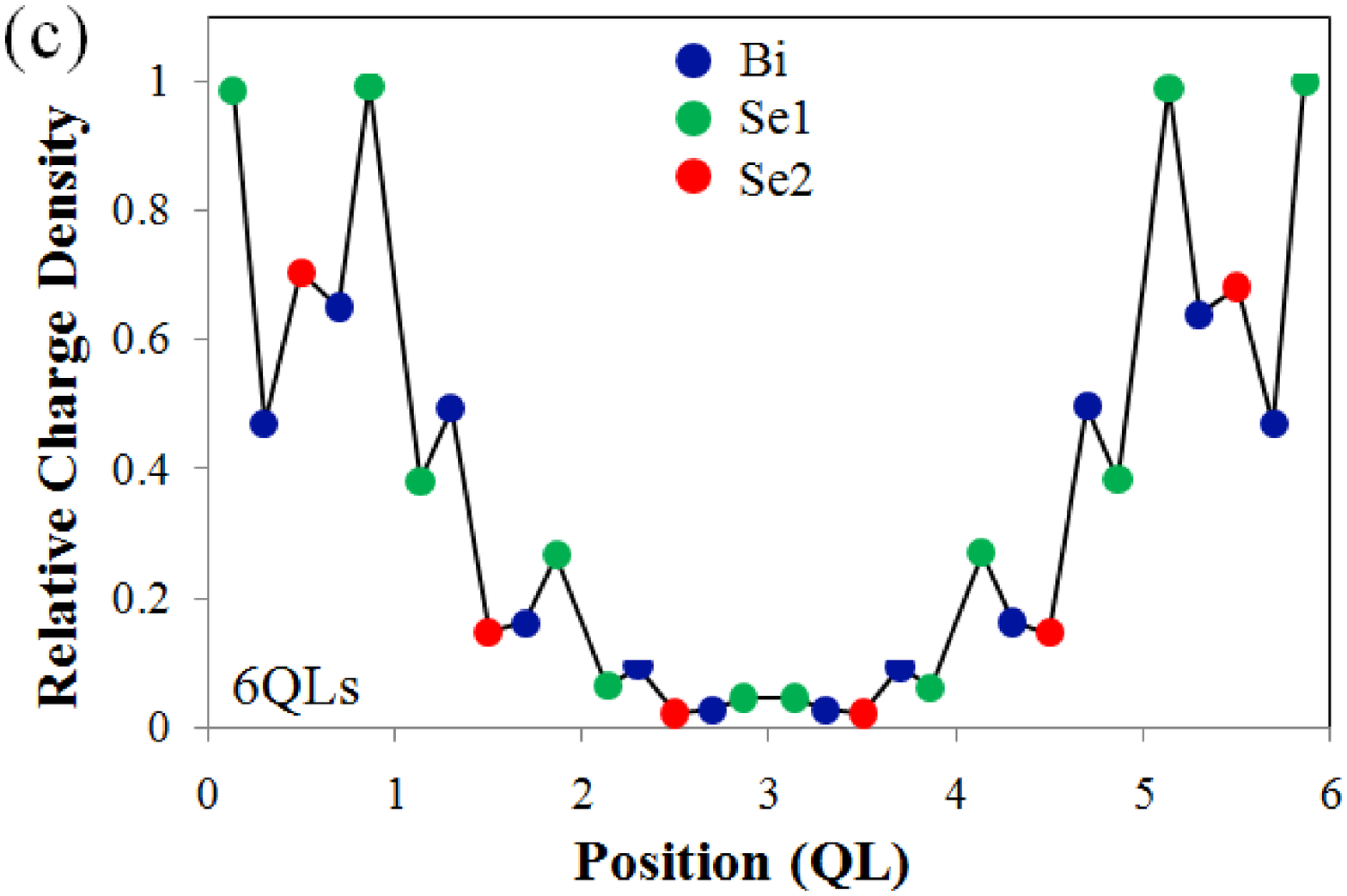}}
\caption{ (Color online) Layer projected relative charge density, contributed by surface state in the neighborhood of the $\bar{\Gamma}$-point in the energy range of 50 meV, for three different thicknesses of QLs (a) 4QLs, (b) 5QLs and (c) 6QLs. For 3QLs and 2QLs no such layer projected charge density could be obtained because of coexistence of bulk as well surface states in the chosen energy range. The color labelling scheme is consistent with that in Figure 1(a).}
\label{fig:Fig5}
\end{figure}

\begin{figure}
\scalebox{0.27}{\includegraphics[angle=-90]{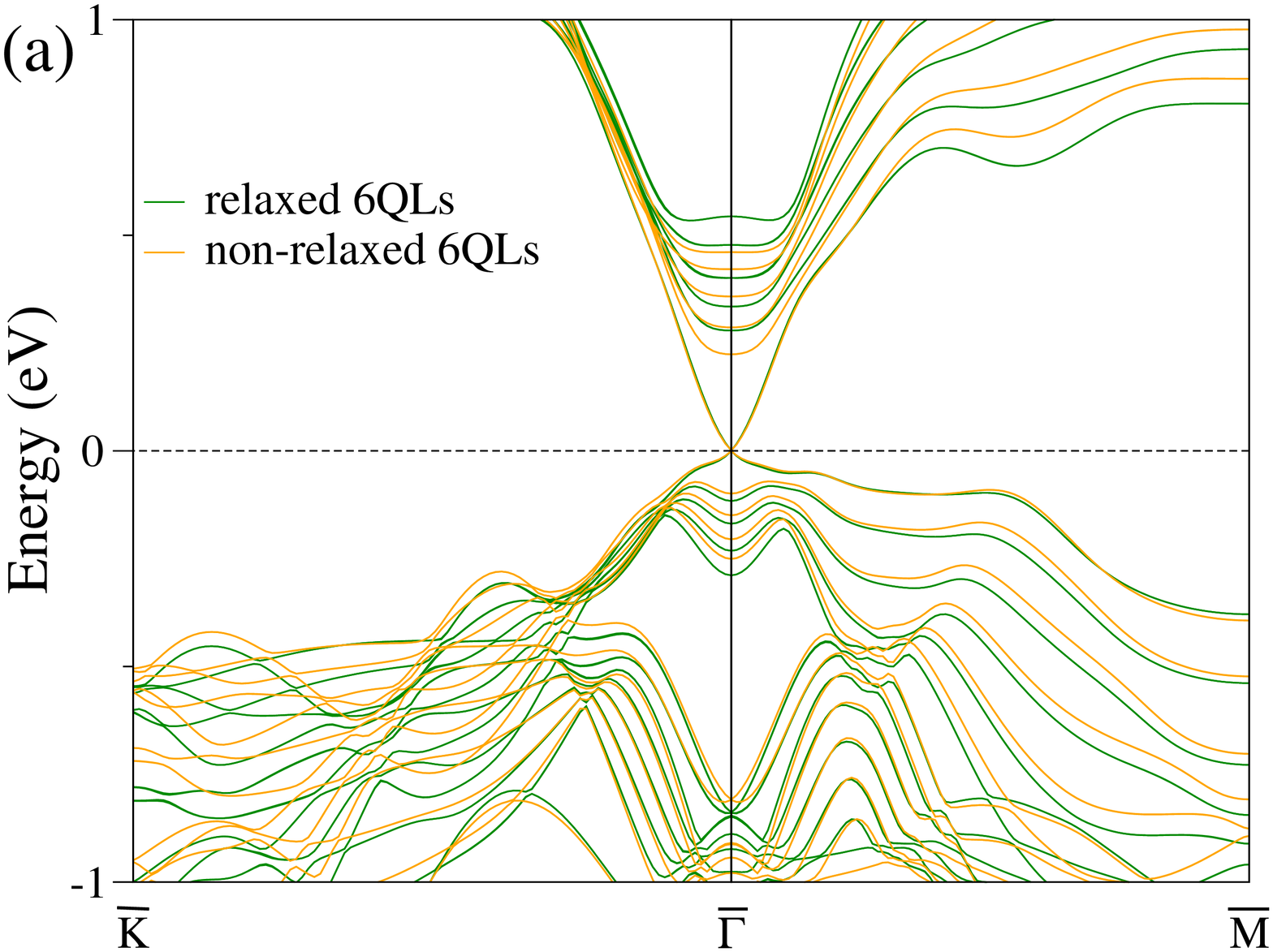}}\\
\scalebox{0.27}{\includegraphics[angle=-90]{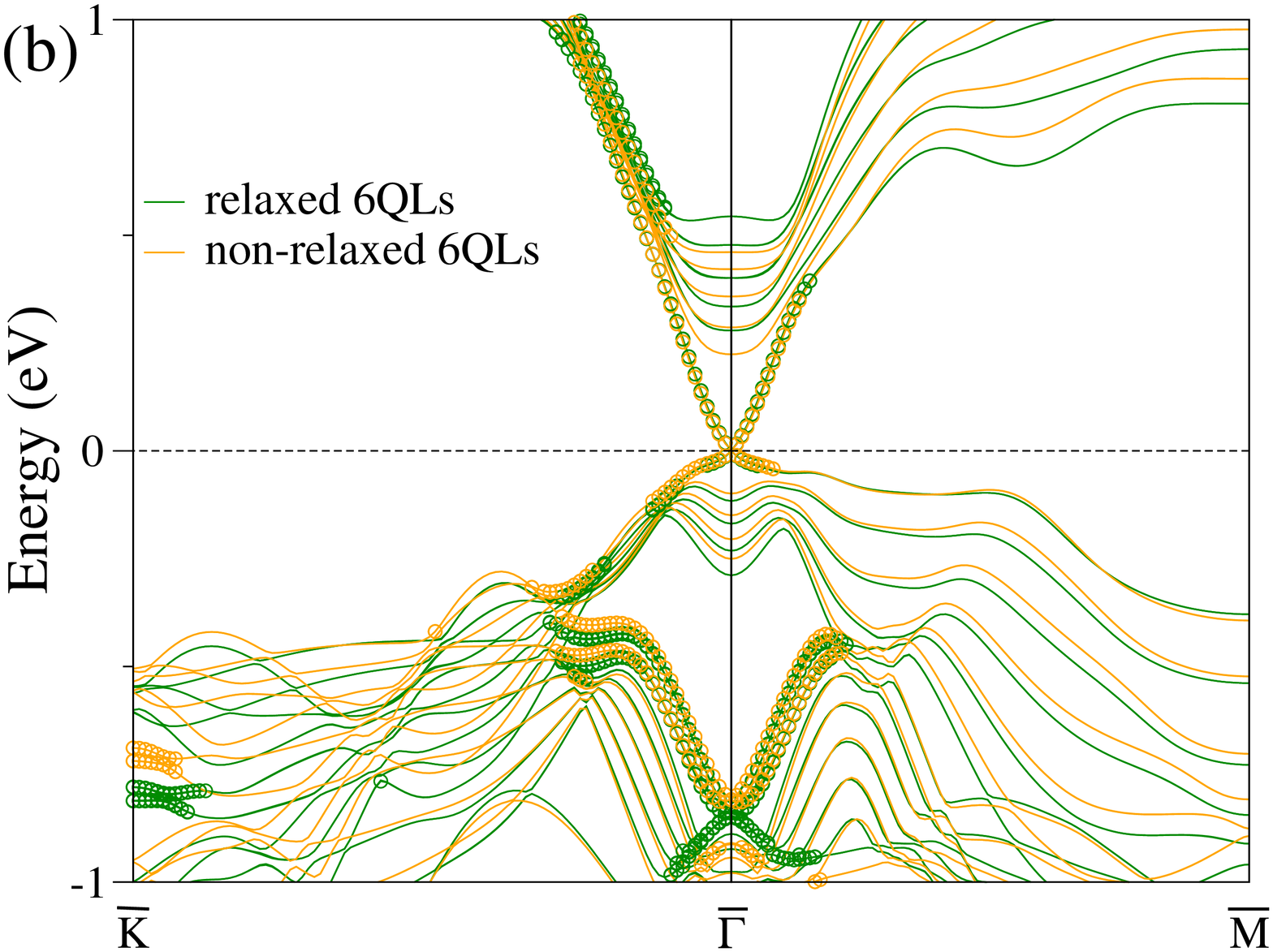}}\\
\scalebox{0.37}{\includegraphics[angle=0]{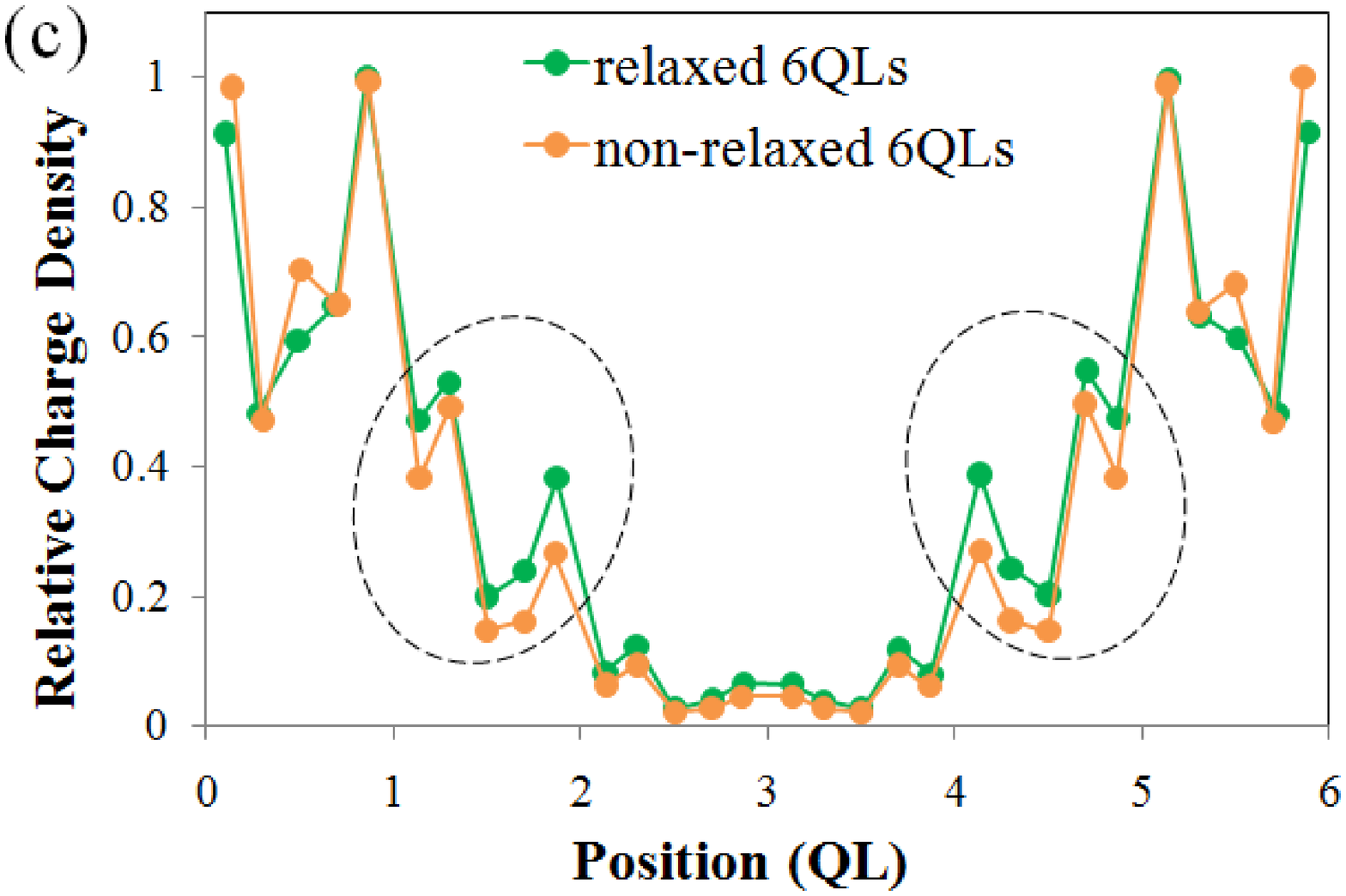}}
\caption{ (Color online) Effects of atomic relaxations on the band structure and charge density distribution of surface states in 6QLs thick thin-film (a) band structure with and without atomic relaxations (b) band structure with orbital contributions 
marked with circles and (c) relative charge density distribution among the atomic layers. The circled region indicate the accumulation of charges with atomic relaxations.}
\label{fig:Fig6}
\end{figure}

\section{Effects of Atomic Relaxations and the Dielectric films}

As the Bi$_2$Se$_3$ thin film is contructed from the bulk crystal, atomic relaxations are inevitable. As a result, both Dirac cone degeneracy and surface state localization length may be affected. Since the 6QLs electronic spectrum has no gap, it is natural to consider this structure for the relaxation study. We considered the same computational parameter as the unrelaxed case but allowed only the {\it z}-component of the atom to relax. Since a vacuum along {\it z}-direction is present, atomic motions are more likely near the vacuum region than within the rigid planar structure. The total energy is assumed to have converged when the {\it z}-component of the Hellman-Feynman force is smaller than 0.025 eV/\AA. We find that the Dirac cone degneracy is not perturbed with atomic relaxations (Fig. 6(a)) and overall band structure features are retained except for small rearrangements of bands throughout the spectrum. This result suggests the robustness of the topologically protected surface states with respect to atomic relaxations. The extraction of surface states for each band and each {\bf k}-point is performed as explained in Section III, and the results are denoted by circles (Fig. 6(b)). The charge density distribution for each layer is determined in a similar way to that in section IV. Figure 6(c) shows the relative charge density of the relaxed structure compared to the non-relaxed case. The overall trends exhibited for the non-relaxed case is maintained, but the charge density seems to be enhanced in the second QL from each side of the surface. This hints at strong spread of the surface states in the second QL as compared to the non-relaxed case.

We considered thin dielectric films of crystalline boron nitride (BN) and quartz (or SiO$_2$) to study their effect on surface states. We note here that amorphous SiO$_2$ is used in practical devices so our results for quartz are at best qualitative. We first discuss the interface structure of BN and Bi$_2$Se$_3$ and its effect on surface band structure of Bi$_2$Se$_3$. BN crystallizes in hexagonal structure with {\it a} = 0.2494 nm and {\it c}= 0.333 nm with d(B-N) =0.144 nm\cite{CrystalBN}. The bulk direct band gap is 5.97 eV\cite{kanda}. We use GGA for Bi$_2$Se$_3$ studies. However, use of GGA to optimize the bulk positions and lattice parameters for BN results in overestimated value of lattice constants and it underestimates the bandgap by 33$\%$ \cite{brink}. Therefore, for the interface studies, we used experimental values of bulk lattice parameters and positions. With these experimental values, our calculated bulk band gap value is same as fully optimized DFT calculations suggest\cite{openmx}.

\begin{figure}[!]
\scalebox{0.55}{\includegraphics{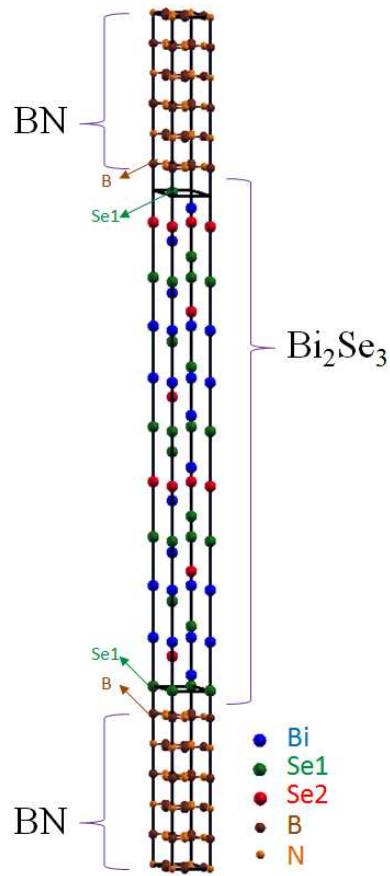}}
\caption{ (Color online) Schematic diagram of supercell structure of BN and Bi$_2$Se$_3$. The colors corresonding to atoms of both materials labled on the right bottom. The arrows on the top indicate the misorientation of B and Se atoms in the interface region on one side of the Bi$_2$Se$_3$ film, whereas the bottom arrows show Se exactly on the top of B. In our DFT calculations, therefore, we are restricted to consider only one interface region between the TI and BN to maintain the orientation. The TI region consists of six QLs of Bi$_2$Se$_3$ and six layers of B and N.}
\label{fig:Fig7}
\end{figure}

\begin{figure}[!]
\scalebox{0.55}{\includegraphics{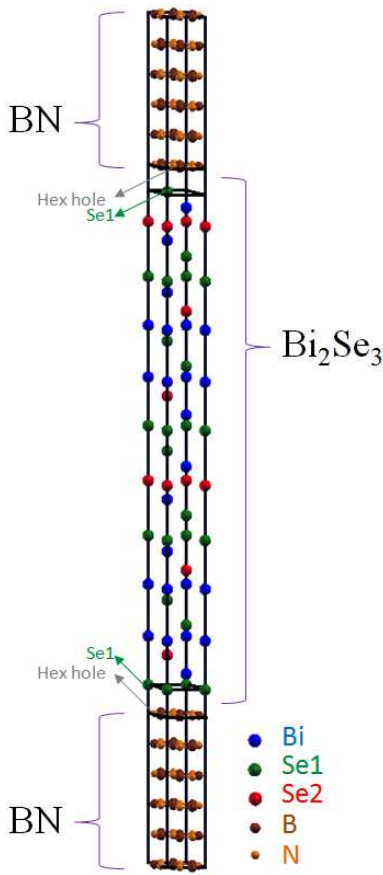}}
\caption{ (Color online) Same as Figure 7 but Se is now on the top of hexagonal hole of BN lattice indicated by arrows. For symmetry regions, we are not restricted here by only one interface region between TI and BN.}
\label{fig:Fig8}
\end{figure}

\begin{figure}[!]
\scalebox{0.27}{\includegraphics[angle=-90]{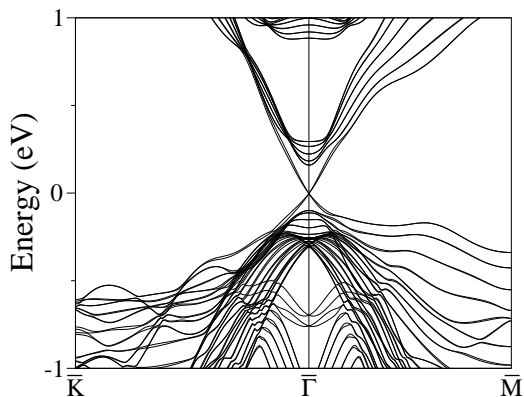}}
\caption{Band structure of Bi$_{2}$Se$_{3}$/BN supercell along high symmetry directions. The Dirac cone and its degeneracy are not disturbed in presence of thin film of crystalline BN. The bulk band gap of Bi$_2$Se$_3$ is reduced by a negligible amount.
}
\label{fig:Fig9}
\end{figure}

Our analysis suggests that to keep our computational burden in DFT-based calculations minimal, 3 $\times$ d(B-N) lattice structure can be matched in-plane with 1 $\times$ 1 Bi$_2$Se$_3$ cell or any multiple of this combination that can maintains 3:1 lattice size ratio resulting always in 4$\%$ natural strain. Any other combinations result in either difficulty in forming the periodic structure or large supercell-size with increase of number of atoms by orders of magnitude to allow strain less than 4$\%$. Along the stacking direction, 6 QLs of Bi$_2$Se$_3$ ($\sim$ 6 nm), which has no band gap for surface states, is put on six layers of BN ($\sim$ 1.7 nm). The choice of 6 BN layer is somewhat arbitrary and guided by the fact the size of vacuum and BN layers should be thick enough to avoid interactions of periodically repeated Bi$_2$Se$_3$ surface layers. Since the atomic displacements in Bi$_2$Se$_3$ play no role in breaking the surface state degeneracy, we strained the BN lattice, instead of the TI lattice, to fit with Bi$_2$Se$_3$ surface resulting in compressive strain. We did not relax the interfacial atomic positions to check whether the strained BN lattice can affect the Dirac cone at the $\Gamma$ point. We considered four configurations of Se positions with respect to boron and nitrogen positions in BN: Se on the top of B, on the top of N, on the hexagonal hole and in the interstitial region between B and N. We note here that BN layers cannot be put on both sides of Bi$_2$Se$_3$ film for Se on the top of B configuration because the same orientations are not maintained between the positions of Se and B (Fig. 7). This occurs because of need to satisfy simultaneously the symmetry of Bi$_2$Se$_3$ layers and BN layers in forming the supercell structure. The same holds true for Se on the top of N atom. However, for Se on the top of the hexagonal hole or at the interstitial region between B and N, the orientations are same on both sides (Fig. 8). For the sake of consistency, we calculated the band structures of the interface structures with BN only on one side of Bi$_2$Se$_3$ for all configurations. For the interface simulations, we used 7 $\times$ 7 $\times$ 1 {\bf k}-point mesh and 2 nm thick vacuum layer. We adjusted the separation between the Se layer and the BN layer. We found the optimal separation of 0.3 nm, and all the studied configurations of Se with respect to BN layer are energetically close and give quite similar band structures. We chose one of these configurations namely Se on B for further discussions. Figure 9 shows the band structure of Bi$_2$Se$_3$ on BN at the optimal interfacial separation of 0.3 nm. It is evident that BN has negligible effect on the surface state dispersion of Bi$_2$Se$_3$ and the Dirac cone degeneracy at the $\Gamma$-point is preserved with no change in the band gap of BN layers. However, the bulk band gap value of Bi$_2$Se$_3$ in the slab structure changes slightly.

We now discuss our results for surface state dispersion in presence of crystalline quartz (or $\alpha$-SiO$_2$). We put quartz on both sides of the TI film. The hexagonal crystal structure of quartz contains four-fold coordinated oxygens, forming a layered structure with Si. The lattice parameters are {\it a}=0.4914 nm and {\it c}=0.5408 nm\cite{quartz}. Our DFT calculations of optimized lattice paramters with PBE potentials are found to be close to the experimental values, accurate to within 0.1$\%$. Therefore, in this section we take experimental lattice parameters for building our interface structure of TI and quartz. The bulk amorphous quartz has direct band gap of 8.9 eV\cite{sio2gap} and our DFT calculation on crystalline SiO$_2$ results in a bandgap value of 6 eV. The interface supercell structure consists of six quintuple layers of Bi$_2$Se$_3$ and two unit cells of SiO$_2$ stacked along the {\it z}-axis. The size of the SiO$_2$ and the Bi$_2$Se$_3$ cell along {\it x}-{\it y} direction chosen is, respectively, 1 $\times$ 1 and 2 $\times$ 2 to keep the computational burden minimal. This produces about 2.75$\%$ compressive strain on both sides of the TI surface. Consideration of larger sizes results in lower strain but the total number of atoms in the cell increases atleast an order of magnitude (250 versus 2500). We did not relax the interfacial atomic positions to check whether the resulting strain can affect the TI Dirac cone. We use 7 $\times$ 7 $\times$ 1 {\bf k}-point mesh for the BZ integration. Periodicity is assumed along {\it x}- {\it y} direction. We performed electronic structure calculations at various interplaner distances between the quartz and Bi$_2$Se$_3$ to fix the optimal distance at which the dielectric oxide effects can be assessed. We consider two surface terminations for quartz: Si- and O-termination each with and without hydrogen passivation of dangling states (Fig. 10). Without hydrogen saturation in O-terminated quartz, the Dirac cone of TI is destroyed whereas in Si-terminated quartz, it is preserved. The interplaner separation at the interface region is found to be  0.3 nm for Si-terimated quartz without passivation. Hydrogen saturation reduces the optimal separation to 0.25 nm in O-termination but the optimal distance is insenstitive to the hydrogen saturation in Si-termination. Figures 11  show band structures of the TI in presence of oxygen terminated quartz with hydrogen passivation. Similar band structure is obtained with Si-termination (Figure not shown). The Dirac cone is seen to be insensitive to both the terminations with hydrogen saturation. The band gaps of thin films of quartz with hydrogen saturated oxygen and with Si terminations are calculated to be 9.23 and 8 eV, respectively. In presence of TI, gaps of each of these terminations reduce but the bulk gaps of TI remain unchanged.                

\begin{figure}[!]
\scalebox{0.50}{\includegraphics{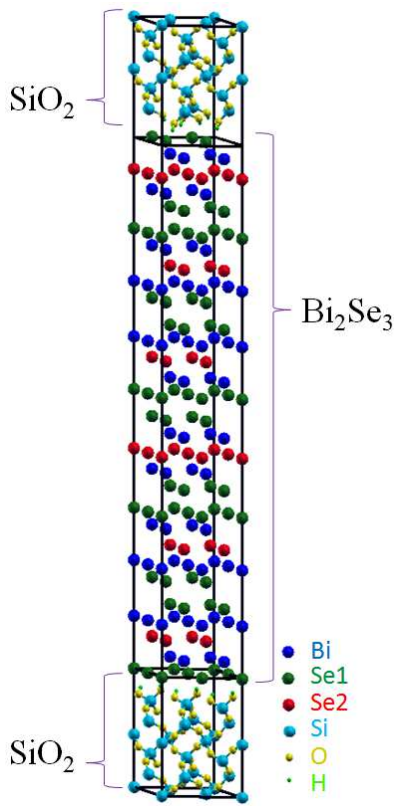}}
\caption{ (Color online) Schematic diagram of oxygen terminated crystalline quartz with dangling states saturated with hydrogen atoms (small green circles) interfaced on both sides of the TI surface. The TI consists of six QLs of Bi$_2$Se$_3$ and two unit cells of quartz.}
\label{fig:Fig10}
\end{figure}

\begin{figure}[!]
\scalebox{0.27}{\includegraphics[angle=-90]{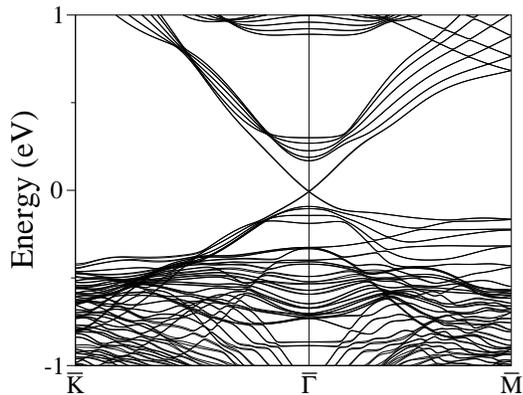}}
\caption{ Band structure of Bi$_{2}$Se$_{3}$ and oxygen terminated quartz passivated with hydrogen atoms along high symmetry directions. The Dirac cone of TI is preserved in this case as well as in Si-terminated quartz.}
\label{fig:Fig11}
\end{figure}

\section {Summary and Conclusions}
We use a density functional based electronic structure method to study the effect of intrinsic and extrinsic perturbations on
the linear spectrum of a strong topological insulator Bi$_2$Se$_3$. Narrow Bi$_2$Se$_3$ film thickness, atomic relaxation, and applied dielectric films of BN and quartz are considered as perturbations. The thickness of the Bi$_2$Se$_3$ film has a considerable effect on the surface state Dirac cone, inducing a gap due to overlap of surface states orginating from two sides of the film. We estimated surface state localization length, by determining the layer projected valence charge density from the states with energies spanning few tens of meV around the Fermi level. The localization length is found to be within 2-3QLs from each side of the film. With increasing overlap between the surface states in thinner structures, the induced gap increases monotonically. We map out the atom and orbital projected bands from the crystal wave-functions. We then extract the surface state contributions to the thin-film band structure and explain the band inversion in bulk TI when spin-orbit interaction is included. Our study hints at insensitivity of the Dirac cone to both type of perturbations. We hope that our studies will promote studies of interplay between surface states and other perturbations.        

\acknowledgments
The authors acknowledge financial support from SWAN-NRI. We thank Texas advanced computing center (TACC) for computational support.

\end{document}